\documentclass[twocolumn,superscriptaddress,aps]{revtex4-2}
\usepackage{hyperref}
\usepackage{graphicx}% Include figure files
\usepackage{floatrow}
\usepackage{subfig}
\usepackage{amsmath}
\usepackage{mathtools}
\usepackage{physics}
\usepackage{amssymb}
\usepackage{dcolumn}% Align table columns on decimal point
\usepackage{bm}% bold math
\usepackage{physics}
\usepackage{xcolor}

\usepackage{ragged2e}
\DeclareCaptionJustification{JustifyCaption}{\justifying}
\begin{document}

%Title of paper
\title{Two-mode Open Quantum Systems:\\Decoherence and Localized Bound State Dynamics}

\author{Chia-Yi Lin}
\author{Chuan-Zhe Yao}
\author{Hon-Lam Lai}
\author{Chin-Chun Tsai}
\author{Wei-Min Zhang}
\email[]{wzhang@mail.ncku.edu.tw}

\affiliation{Department of Physics and Center for Quantum Information Science, National Cheng Kung University, Tainan 70101, Taiwan}
%Collaboration name if desired (requires use of superscriptaddress
%option in \documentclass). \noaffiliation is required (may also be
%used with the \author command).
%\collaboration can be followed by \email, \homepage, \thanks as well.
%\collaboration{}
%\noaffiliation

\date{\today}

\begin{abstract}
Dissipationless localized bound states of open quantum systems are significantly robust to decoherence and have potential applications in quantum technologies. In this work, the decoherence dynamics and dissipationless localized bound states of a two-mode open quantum system are investigated. The conditions for the emergence of dissipationless localized bound states are analytically solved, and the corresponding critical system-environment couplings under different values of the inter-mode coupling and the detuning are determined. The decoherence dynamics of the system under such conditions are analyzed and dissipationless coherence between the different localized bound states against decoherence is clearly shown. This may provide a new avenue to develop dissipationless quantum technology for quantum operations.
\end{abstract}

% insert suggested keywords - APS authors don't need to do this
%\keywords{}
%\maketitle must follow title, authors, abstract, and keywords
\maketitle

% body of paper here - Use proper section commands
% References should be done using the \cite, \ref, and \label commands
\section{Introduction} 
Quantum information processing holds promise for computation, communication, and cryptography by utilizing the unique properties of quantum mechanics, namely the quantum coherence and quantum entanglement \cite{NielsenChuang2010}. However, practical implementation of quantum information protocols faces significant challenges, one of the most notable being the decoherence. Decoherence arises from the interaction of quantum systems with their surrounding environments, leading to the loss of quantum coherence. Therefore, overcoming decoherence is essential for realizing the full potential of quantum technologies. For this purpose, quantum systems must be treated as open systems and different approximated master equations for the description of decoherence dynamics have been derived, for examples, the Gorini–Kossakowski-Sudarshan-Lindblad master equation \cite{GKS1976, Lindblad1976}, and the Born-Markov master equation as well as the truncated Nakajima-Zwanzig master equation \cite{Nakajima1958, Zwanzig1960, BP2007} under various approximations. The approximations in these master equations cause the system-environment back actions to be overly simplified. Consequently, the dissipation and fluctuation effects induced by system-environment coupling cannot be correctly manifested and understood. 
%, which is associated with many-body localization \cite{Xiong2015}.

In the past two decades, by extending the Feynman-Vernon influence functional approach \cite{Feynman1963} to the coherent state path integral approach \cite{WMZ1990}, we have developed the exact master equations for various fermionic open systems \cite{Tu2008,TLZ2009,Jin2010,WMZ2012,WMZ2019}, bosonic open systems \cite{Lei2012,Xiong2010,wu2010non,YZ2022}, as well as for hybrid topological systems \cite{Lai2018,Yao2020, YW2020,Yao2023}. The previous achievement on exact master equation using the Feynman-Vernon influence functional approach was given only for quantum Brownian motion \cite{Leggett1983,HPZ1992}. However, as we have recently shown \cite{YZ2022,YZ2024}, the master equation for quantum Brownian motion obtained in \cite{Leggett1983,HPZ1992} mis-separate a part of the energy renormalization from the dissipation dynamics. In fact, the dissipation and fluctuation dynamics of open systems can be depicted uniquely and unambiguously through the non-equilibrium Green functions we introduced for open quantum systems \cite{Jin2010, Lei2012,WMZ2012, WMZ2019,YW2020}, which include all the non-Markovian effects induced by the system-environment backactions. Specifically, for strong system-environment couplings, dissipationless localized bound states are emerged in open quantum systems as a long-time non-Markovian quantum memory dynamics \cite{WMZ2012,wu2010non,Xiong2015,Lu2015,Lai2018}. Such long-time non-Markovian memory dynamics is fully against from the decoherence and thermalization \cite{WMH2022}. In the literature, extensive attentions have been focused on topological quantum 
computing based on Majorana zero mode bound states because it has been thought to be robust against decoherence \cite{Kitaev03,Alicea11,Fu15,Plugge16,Lutchyn18}. However, the long-time search of the Majorana bound states has not been successful in experiments, while theoretically we have proven that Majorana bound states (if exist) still suffer from the same decoherence problem as other qubit systems for qubit manipulations \cite{Lai2018,Yao2020,YW2020,Yao2023}. Thus,  the dissipationless localized bound states in open quantum systems may offer a new avenue towards achieving dissipationless (truly decoherence-free) quantum computation in the presence of environmental noises and the engineering environmental disturbances. 

In exploring the possibility of utilizing these localized bound states in quantum technologies, one of the crucial elements is the controllability of the the localized bound states under various tunable parameters of the system and its environments. In our previous works \cite{wu2010non,Xiong2010,Xiong2015,Lu2015,WMH2022}, we showed that the localized bound states can be formed by tuning the system-environment couplings in a single-mode system. In this paper, we further consider the localized bound states of an open system with two bosonic modes, with tunable inter-mode coupling and system-environment couplings. We find that, in this two-mode system, localized bound states can be controlled not only by the system-environment couplings, but also by the inter-mode coupling and the detuning of the system. More importantly, rather than the single localized bound state in single-mode system, the conditions for emerging multiple localized bound states are provided in this two-mode open system. When two localized bound states are formed, the coherence between the two  localized bound states also exhibit dissipationless with controllable frequencies. These phenomena present the potential for realizing dissipationless quantum computation, where the qubits and qubit manipulations are built with dissipationless localized bound states.

The rest of the paper is arranged as follows. In Sec.~II, we introduce the two-mode system coupled to environments and determine the conditions for the formation of the localized bound states. In Sec.~III, we explicitly show the conditions for the existence of the localized-mode state  under Ohmic-type spectral density, and determine the critical system-environment couplings at which the localized bound states emerge. Moreover, we show that when there are two localized bound states, the system exhibits dissipationless oscillations (coherence), of which the frequency can be tuned by various system parameters. Finally, a conclusion is given in Sec.~IV. 

\section{Model and formalism}
We consider a bosonic system with two modes respectively coupled to two environments. These two modes are further coupled to each other. A schematic plot of the system is presented in Fig.~1. The Hamiltonian of the total system can be described by  
\begin{align}
	H_{tot}  = & ~\hbar\omega_{s, 1}{a_1^{\dagger}}{a_1} + \hbar\omega_{s, 2}{a_2^{\dagger}}{a_2} + \hbar \kappa\left({a_1^{\dagger}}{a_2} + {a_2^{\dagger}}{a_1}\right) \notag\\ 
	& + \sum_{k}\left(\hbar\omega_{L k}{b_{L k}^{\dagger}}{b_{L k}} + \hbar\omega_{R k}{b_{R k}^{\dagger}}{b_{R k}}\right) \notag\\
	& + \hbar\sum_{k}\left(W_{1Lk}{a_{1}^{\dagger}}{b_{Lk}} + W_{2Rk}{a_{2}^{\dagger}}{b_{Rk}} + \text{H.c.}\right), \label{2MQD}
\end{align}
where $a_{i}^{\dagger}$($a_{i}$) and $b_{\alpha k}^{\dagger}$($b_{\alpha k}$) are the bosonic creation (annihilation) operators of the system and environment respectively, with frequency $\omega_{s, i}$ ($i = 1, 2$) and $\omega_{\alpha k}$ ($\alpha = L, R$). Here, $\kappa$ is the coupling strength between the two modes. The last term of Hamiltonian describes the system-environment interaction, where the parameter $W_{i\alpha k}$ is the coupling strength between the system mode $i$ and the environment mode $k$ of $\alpha$. Both inter-mode couplings and system-environment couplings are tunable experimentally, for instance, the beam splitters in integrated photonic circuit experiments  \cite{QIC,Paesani2019,Arrazola2021}.

\begin{figure}
    \centering
    \includegraphics[width=0.85\linewidth]{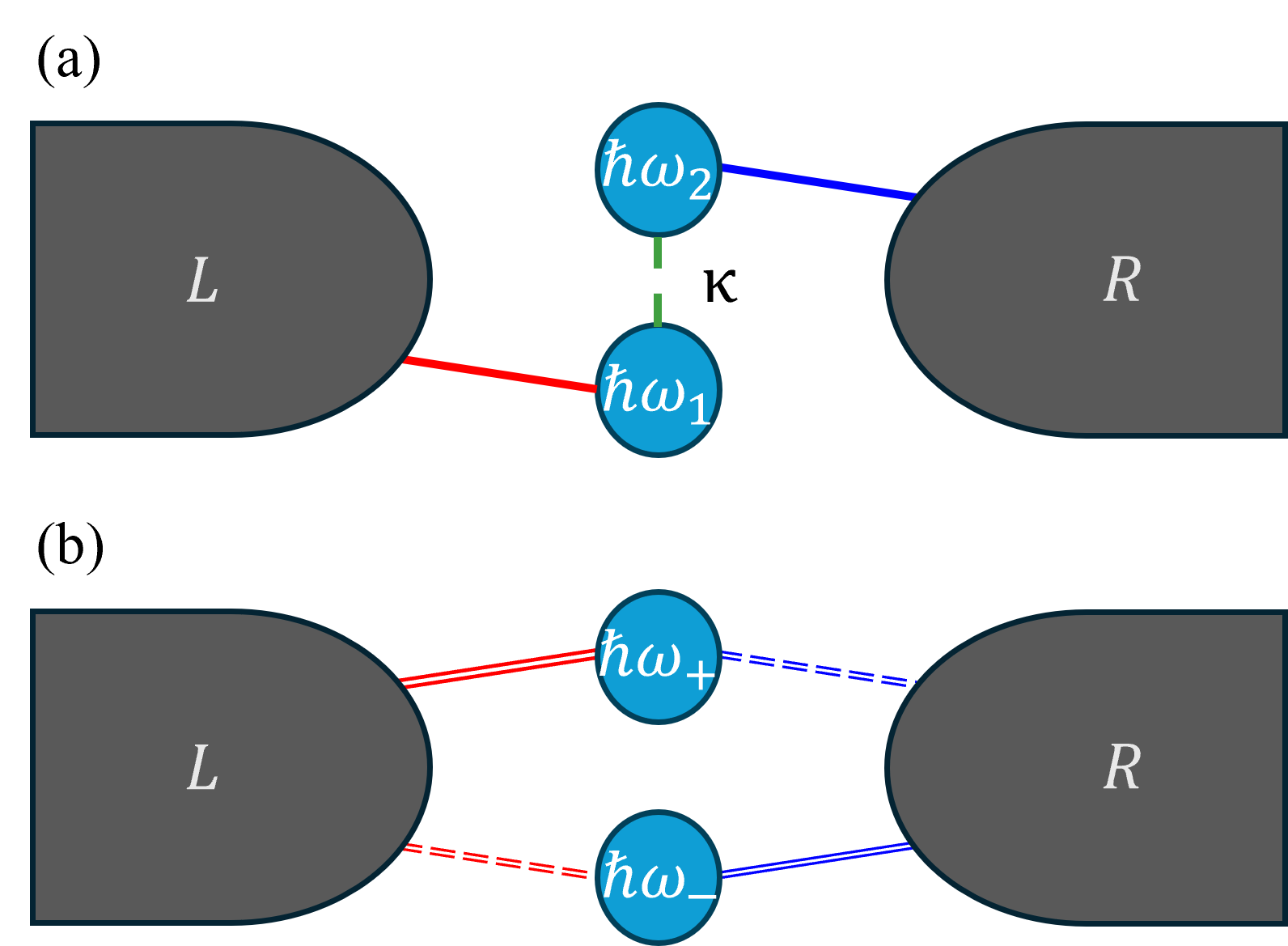}
    \caption{(Color online) A schematic diagram of the two-mode open quantum system coupled to two reservoirs $\alpha = L, R$ (a) in the basis of the two modes described by Eq.~(\ref{2MQD}) and (b) in the new basis of two eigenmodes in terms of the transformation Eq.~(\ref{modestransform}) described by Eq.~(\ref{2MQD_dg}).} 
    \label{A schematic diagram}
\end{figure}

Based on the Hamiltonian (\ref{2MQD}), the reduced density matrix of the two-mode system obeys the following exact master equation \cite{Lei2012, WMZ2012, WMZ2019},
\begin{align}
	\frac{d\rho(t)}{dt}=&~\frac{1}{i}[\tilde{H}_{S}(t),\rho(t)] \notag \\
	& +\sum_{i,j} \bm{\gamma}_{ij}(t)[2a_{j}\rho(t)a^{\dag}_{i} -a^{\dag}_{i}a_{j}\rho(t)-\rho(t)a^{\dag}_{i}a_{j}] \notag\\
    & + \sum_{i,j} \tilde{\bm{\gamma}}_{ij}(t)[a^{\dag}_{i}\rho(t)a_{j} +a_{j}\rho(t)a^{\dag}_{i}-a^{\dag}_{i}a_{j}\rho(t) \notag \\
    & \qquad  \qquad  \quad -\rho(t)a_{j}a^{\dag}_{i}], \label{EOM}
\end{align}
where $i,j=\{1,2\}$. The first term of the above master equation describes the system unitary evolution with the renormalized Hamiltonian $\tilde{H}_{S}(t)=\sum_{ij} \hbar \tilde{\bm{\omega}}_{sij}a^{\dag}_{i}a_{j}$. The second and third terms describe the dissipation and fluctuation dynamics, respectively. The renormalized frequencies of the system $\tilde{\bm{\omega}}_{s}(t)$, the dissipation coefficient $\bm{\gamma}(t)$ and the fluctuation coefficient $\tilde{\bm{\gamma}}(t)$ are given by
\begin{subequations}
\begin{align}
    \tilde{\bm{\omega}}_{s}(t)&=\frac{i}{2}\Big[\dot{\bm{U}}(t, t_0)\bm{U}^{-1}(t, t_0)-{\rm H.c.}\Big], \\
    \bm{\gamma}(t)&=-\frac{1}{2}\Big[\dot{\bm{U}}(t, t_0)\bm{U}^{-1}(t, t_0)+{\rm H.c.}\Big], \\
    \tilde{\bm{\gamma}}(t)&=\bm{V}(t,t)-\Big[\dot{\bm{U}}(t, t_0)\bm{U}^{-1}(t, t_0)\bm{V}(t,t)+{\rm H.c.}\Big].
\end{align}
\end{subequations}
Here the $2\times 2$ matrices $\bm{U}(\tau, t_0)$  and $\bm{V}(\tau, t)$ are the non-equilibrium Green functions of open systems
we introduced in our previous derivation \cite{Jin2010,Lei2012,WMZ2012,WMZ2019} and obey the following integro-differential equations: 
\begin{subequations}
\label{UVT}
	\begin{align}
		\frac{d}{d\tau} \bm{U}(\tau, t_0) & + i\bm{\omega_s}\bm{U}(\tau, t_0) \nonumber \\ 
		& + \int_{t_0}^{\tau}d\tau'\bm{G}(\tau, \tau')\bm{U}(\tau', t_0) = 0, \label{integroU}
		\\
		\frac{d}{d\tau} \bm{V}(\tau, t) & +  i\bm{\omega_s}\bm{V}(\tau, t) + \int_{t_0}^{\tau}d\tau'\bm{G}(\tau, \tau')\bm{V}(\tau', t) \nonumber \\ 
		= & \int_{t_0}^{t}   d\tau'\bm{\tilde{G}}(\tau, \tau'){\bm{U}^{\dagger}}(\tau', t) \quad (t_0 \leq \tau \leq t), \label{integroV}
	\end{align}
\end{subequations}
subjected to the boundary condition $\bm{U}(t_0, t_0) = \bm{I}$ and $\bm{V}(t_0, t) = \bm{0}$, where $\bm{\omega_s}$ is the frequency matrix of the system given in Eq.~(\ref{2MQD}). Moreover, Eq.~(\ref{integroV}) can be solved analytically in terms of $\bm{U}(\tau, t_0)$:
\begin{align}
	\bm{V}(\tau, t) = \int_{t_0}^{\tau}{dt_1}\int_{t_0}^{t}{dt_2}\bm{U}(\tau, t_1)\bm{\tilde{G}}(t_1, t_2)\bm{U}^{\dagger}(t_2, t),
\end{align}
which gives the general noneqilibrium dissipation-fluctuation theorem for open quantum systems  \cite{Jin2010,Lei2012,WMZ2012,WMZ2019}.

By defining the spectral density of the environment $J_{\alpha ij}(\omega) = 2\pi\sum_{k}W_{i\alpha k}W_{j\alpha k}^{*}\delta(\omega-\omega_{\alpha k})$, the integral kernels in Eq.~(\ref{UVT}) can be expressed as
\begin{subequations}
	\begin{align}
		\bm{G}_{ ij}(\tau, \tau') & = \sum_\alpha \int\frac{d\omega}{2\pi}J_{\alpha ij}(\omega)e^{-i\omega(\tau-\tau')}, \\
		\widetilde{\bm{G}}_{ ij}(\tau, \tau') & = \sum_\alpha \int\frac{d\omega}{2\pi}J_{\alpha ij}(\omega)f_\alpha (\omega)e^{-i\omega(\tau-\tau')}.
	\end{align}
\end{subequations}
Here, we parameter the different spectral density matrix elements, which corresponds to the two modes coupled separately to two environments with the same spectral structure, namely, %$\omega_{\alpha k} = \omega_k$, and
\begin{subequations}
    \begin{align}
        & J_{L11} (\omega)= J_{L} (\omega)= \lambda J(\omega), \\
        & J_{R22}(\omega) = J_{R}(\omega) =  \left(2-\lambda\right)J(\omega),
    \end{align} 
\end{subequations}
with the constraint parameter $1 \geq \lambda \geq 0$. The function $f_\alpha (\omega) = 1/(e^{\hbar\omega /k_BT_0}-1)$ is the initial particle distribution given by the Bose-Einstein distribution for both environments being the thermal state with the same temperature $T_0$ at the initial time $t_0$. 

Apply the modified Laplace transformation $\widetilde{\bm{U}}(z) = \int_{t_0}^{\infty}dt\bm{U}(t, t_0)e^{iz(t-t_0)}$ to Eq.~(\ref{integroU}) \cite{WMZ2012}, one has
\begin{align}
	\widetilde{\bm{U}}(z) 
	& = i\left(z\bm{I} - \bm{\omega}_s - \bm{\Sigma}(z)\right)^{-1}  \notag \\
	& = i\left(\begin{array}{cc}
		z-\omega_{1}-\Sigma_{L}(z) & -\kappa \\
		-\kappa & z-\omega_{2}-\Sigma_{R}(z)
	\end{array}\right)^{-1},
    \label{LTU}
\end{align}
where $\Sigma_{\alpha}(z)$ is the self-energy correction which is the Laplace transformation of the integral kernel in Eq.(\ref{integroU}), 
\begin{align}
	\Sigma_{\alpha}(z) = \int\frac{d\omega}{2\pi}\frac{J_{\alpha}(\omega)}{z-\omega} 
	\stackrel{z \pm i0^{+}}{\longrightarrow} \Delta_{\alpha}(z) \mp \frac{i}{2}J_{\alpha}(z),  \label{self_e}
\end{align}
and $\Delta_{\alpha}(z) = \mathcal{P} \int\frac{d\omega}{2\pi}\frac{J_{\alpha}(\omega)}{z-\omega}$ is the principal value of the integral, which induces the frequency (energy) shift of the system due to the coupling to their environment. The general solution of $\bm{U}(t, t_0)$ can then be expressed as:
\begin{align}
    \bm{U}(t, t_0) = & \sum_{\omega_l}{\bm{\mathcal{Z}}}(\omega_l)e^{-iz_l(t-t_0)} + \int \frac{d\omega}{2\pi} \big[\widetilde{\bm{U}}(\omega+i0^{+}) \nonumber \\ 
    & \qquad \qquad - \widetilde{\bm{U}}(\omega-i0^{+})\big] e^{-i\omega(t-t_0)}.   \label{analyticU}
\end{align}
The first term is the contribution from dissipationless localized bound states of the system  with amplitude $\bm{\mathcal{Z}}(\omega_l)$ and frequencies $\omega_l$. It arises from the relatively strong coupling with the environments or the band gaps in the engineering structural environments, and describes the long-time (never damping) non-Markovian dynamics of the system, protecting the system from decoherence \cite{WMZ2012}. The second term is an non-exponential decay arisen from the discontinuity in the imaginary part of the self-energy corrections, as shown in Eq.~(\ref{self_e}). 

The existence of localized bound states depending on the simultaneous fulfillment of the conditions \cite{WMZ2012}: 
\begin{align}
    \omega_l - \omega_{0} \mp \sqrt{\Big[\delta+(1-\lambda)\Delta(\omega_l)\Big]^2 + \kappa^2} - \Delta(\omega_l) = 0, \label{LBS}
\end{align}
with the additional requirement $J(\omega_l) = 0$, where $\omega_{0} = \left(\omega_{1}+\omega_{2}\right)/2$ is the average frequency of the two modes and $\delta = \left(\omega_{2}-\omega_{1}\right)/2$ is the half of the detuning. The localized bound state amplitude is given by
\begin{widetext}
\begin{align}
	\bm{\mathcal{Z}}(\omega_l) = \frac{1}{2\big[M_{0}(\omega_{l}) - \Delta'(\omega_l)(\omega_{l} - \omega_0 + (1-\lambda)\delta - \lambda(2-\lambda)\Delta(\omega_{l}))\big]} \left(\begin{array}{cc}
            M_{R}(\omega_{l}) & \kappa \\
            \kappa   & M_{L}(\omega_{l})
        \end{array} \right)
\end{align}
\end{widetext}
where $M_{0}(\omega) = \omega - {\omega}_{0} - \Delta(\omega)$, $M_{L}(\omega) = \omega - {\omega}_{1} - \lambda\Delta(\omega)$, $M_{R}(\omega) = \omega - {\omega}_{2} - (2-\lambda)\Delta(\omega)$, and $\Delta'(\omega)$ is the derivative of $\Delta(\omega)$ with respect to $\omega$. Notably, Eq. (\ref{LBS}) shows two sets of pole condition indicated by the sign $\mp$, where the frequency $\omega_{l}$ is shifted by the inter-mode coupling $\kappa$ and the detuning $\delta$ in two different directions in the frequency domain. A localized bound state exists when the frequency $\omega_{l}$ is located at the spectrum vanishing regimes $J(\omega_l) = 0$, which arises from the frequency shift contributed by the system-environment coupling, the inter-mode coupling, the detuning as well as the modification of the asymmetry environmental structure. 

For a more clear physical picture, we diagonize the system Hamiltonian. The diagonalized eigenmodes are expressed in terms of the original modes $a_1$ and $a_2$ as 
\begin{subequations} \label{modestransform}
\begin{align}
	a_{-} & = \cos(\frac{\theta}{2}) a_1 - \sin(\frac{\theta}{2}) a_2, \\
	a_{+} & = \sin(\frac{\theta}{2}) a_1  +  \cos(\frac{\theta}{2}) a_2, 
\end{align}
\end{subequations}
where $a_{-}^{\dagger}$($a_{-}$) and $a_{+}^{\dagger}$($a_{+}$) are creation (annihilation) operators of the corresponding eigen-frequencies $\omega_{\pm} = \omega_{0} \pm \sqrt{\delta^2 + \kappa^2}$ and the phase $\theta = \tan^{-1}(\frac{\kappa}{\delta})$.
%, where $\omega_{0} = \left(\omega_{1}+\omega_{2}\right)/2$ is the average frequency of the two modes and 
%$\delta = \left(\omega_{2}-\omega_{1}\right)/2$ is the half of detuning. 
Then, the total Hamiltonian can be rewritten as
\begin{align}
	\begin{split}
		H_{tot} & = \hslash\omega_{-}{a_{-}^{\dagger}}{a_{-}} + \hslash\omega_{+}{a_{+}^{\dagger}}{a_{+}} + \sum_{\alpha k}\hslash\omega_{\alpha k}{b_{\alpha k}^{\dagger}}{b_{\alpha k}} \\ 
		& + \hslash\sum_{\alpha k}\left(W_{\alpha k}^{-} {a_{-}^{\dagger}}{b_{\alpha k}} + W_{\alpha k}^{+} {a_{+}^{\dagger}}{b_{\alpha k}} + \text{H.c.}\right),   \label{2MQD_dg}
	\end{split}
\end{align}
with the eigenmode-environment couplings $W_{\alpha k}^{-}$ and $W_{\alpha k}^{+}$ given by the following rotation,
\begin{subequations}
\begin{align}
	& W_{Lk}^{-} = \cos(\frac{\theta}{2})W_{1Lk}, \quad W_{Rk}^{-} = -\sin(\frac{\theta}{2})W_{2Rk},  \\
	& W_{Lk}^{+} = \sin(\frac{\theta}{2})W_{1Lk}, \quad W_{Rk}^{+} = \cos(\frac{\theta}{2})W_{2Rk}. 
\end{align}
\end{subequations}
The above expression highlights how the inter-mode coupling $\kappa$ and the detuning $\delta$ govern the dynamics of the total system. Furthermore, Eq.~(\ref{2MQD_dg}) shows explicitly that after diagonalization, the two modes are decoupled so that the two eigenmodes couple
simultaneously to the environments on the left and right sides. This is schematically illustrated further in Fig. \ref{A schematic diagram}. 
As a result, Eq.~(\ref{LTU}) is diagnosed in the eigenmode basis
%and would be elaborated on subsequent discussions.
\begin{align}
    \begin{split}
	\widetilde{\bm{U}}(z) 
	& = i\left(\begin{array}{cc}
		z-\omega_{-}-\lambda_{-}\Sigma(z) & 0 \\
		0 & z-\omega_{+}-\lambda_{+}\Sigma(z)
	\end{array}\right)^{-1},
    \end{split}
\end{align}
where
\begin{subequations}
\begin{align}
    \lambda_{-} & = \cos^2(\tfrac{\theta}{2})\lambda + \sin^2(\tfrac{\theta}{2})(2-\lambda), \\
    \lambda_{+} & = \sin^2(\tfrac{\theta}{2})\lambda + \cos^2(\tfrac{\theta}{2})(2-\lambda).
\end{align}
\end{subequations}
The general solution of Green functions ${U_{-}}$ and ${U_{+}}$ can then be reduced to
\begin{subequations}
    \begin{align}
         {U_{-}}(t, t_0) = & {\mathcal{Z}}(\omega_{l_-})e^{-i{\omega_{l_-}}(t-t_0)} \nonumber \\ 
         & + \int \frac{d\omega}{2\pi} \frac{\lambda_{-}J(\omega)e^{-i\omega(t-t_0)}}{\left[\omega - \omega_{-} - \lambda_{-}\Delta(\omega)\right]^2 + \lambda_{-}^2J^2(\omega)/4}, \\
        {U_{+}}(t, t_0) = & {\mathcal{Z}}(\omega_{l_+})e^{-i{\omega_{l_+}}(t-t_0)} \nonumber \\ 
        & + \int \frac{d\omega}{2\pi} \frac{\lambda_{+}J(\omega)e^{-i\omega(t-t_0)}}{\left[\omega - \omega_{+} - \lambda_{+}\Delta(\omega)\right]^2 + \lambda_{+}^2J^2(\omega)/4}. 
    \end{align}
\end{subequations}
Then the localized bound state frequencies $\omega_{l_{-}}$ and $\omega_{l_{+}}$ of two eigenmodes are respectively determined by, 
\begin{subequations}
    \begin{align}
        & \omega_{l_{-}} - \omega_{-} - \lambda_{-}\Delta(\omega) = 0 \quad 
	\text{and} \quad J(\omega) = 0, \\
        & \omega_{l_{+}} - \omega_{+} - \lambda_{+}\Delta(\omega) = 0 \quad 
	\text{and} \quad J(\omega) = 0.
    \end{align}
\end{subequations}
From the above solution, it is evident that the eigenmodes $a_{-}$ and $a_{+}$ are linear combinations of $a_{1}$ and $a_{2}$. When a localized bound state appears in one of the eigenmodes, it is expected to simultaneously affect both of the original physical modes.

\section{Numerical Analysis} \label{sec: numerical-analysis}
\subsection{Emergence of localized bound states}
Based on the formulation given in the previous Section, we can now discuss more explicitly the emergence of the dissipationless localized bound states from the general non-Markovian decoherence theory. Consider a general bosonic environment with Ohmic-type spectral density \cite{Leggett1987}:
\begin{align}
	J(\omega) = 2\pi\eta\omega\left(\frac{\omega}{\omega_c}\right)^{s-1}e^{-\frac{\omega}{\omega_c}}   \label{Ohmic-type}
\end{align}
where $\eta$ is dimensionless coupling strength between the system and environment, and $\omega_c$ is the cutoff frequency of the environmental spectrum. The parameter $s$ classifies the environment as sub-Ohmic ($s < 1$), Ohmic ($s = 1$), or super-Ohmic ($s > 1$). Without loss of generality, we firstly consider the case with symmetric system-environment coupling strength, i.e.~$\lambda = 1$. The spectral densities of the environments are then reduced to $J_L(\omega) = J_R(\omega) = J(\omega)$. From Eq.~(\ref{LBS}), the dissipationless localized bound states occur when
\begin{align}
	\omega_{l_\mp} = \left(\omega_0\mp\sqrt{\delta^2 + \kappa^2}\right) + \Delta(\omega_{l_\mp}) < 0 \label{LBS_lambda1}, 
\end{align}
for $\eta\omega_c\Gamma(s) > \left(\omega_0 \mp \sqrt{\delta^2 + \kappa^2}\right)$, where $\Gamma(s) = \int_{0}^{\infty}x^{s-1} e^{-x} dx$ is a gamma function \cite{WMZ2012}. Therefore, the localized bound states $\omega_{l_{\mp}}$ will appear when the system-environment coupling strength $\eta > \eta_{c_{\mp}}(s) = \omega_{\mp}/\omega_c\Gamma(s)$. 

\begin{figure*}[htb]
    \centering
    \includegraphics[width=0.9 \linewidth]{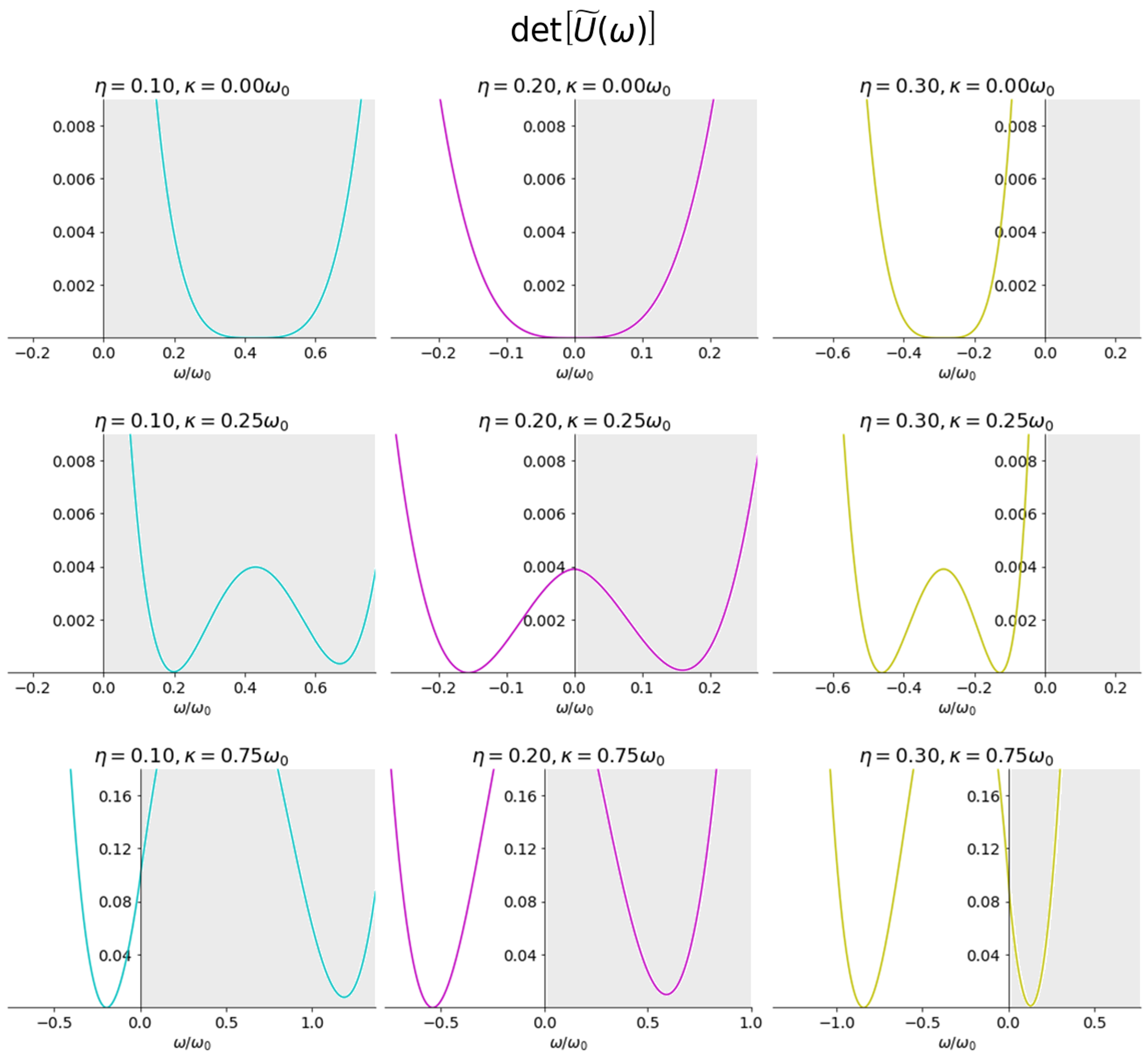}
    \caption{(Color online) The existence of localized bound states in symmetric Ohmic environmental structure ($\lambda = 1$ and $s=1$)  determined by the pole of the $\bm{U}(\omega)$, in the region $J(\omega) = 0$. The plotted lines show the inverse of the determinant ${\rm det}[\widetilde{\bm{U}}(\omega)]$ with different system-environment coupling $\eta$ and inter-mode coupling $\kappa$. Here, we take $\omega_{1}=\omega_{2}=\omega_{0}$, and $\omega_c = 5\omega_0$.}
    \label{UtdiffKap}
\end{figure*}

To understand the emergence of the localized bound states in this two-mode open quantum system, we plot a series of figures to determine the conditions under which the localized bound states emerge. In Fig.~\ref{UtdiffKap}, we present the inverse of the determinant of $\widetilde{\bm{U}}(\omega)$ with different values of the system-environment coupling $\eta$ and the inter-mode coupling $\kappa$ for symmetric couplings ($\lambda=1$) to the Ohmic spectral density ($s=1$). Localized bound states are formed at the pole $\{\omega_{l_{j}}\}$ such that ${\rm det}[\widetilde{\bm{U}}(\omega_{l_{j}})]=0$ with $\omega_{l_{j}}<0$, which contribute to the dissipationless term of Eq.~(\ref{analyticU}). One can see from the plots that when the system-environment coupling $\eta$ is small, there is no zero-crossing point on the negative $\omega$ axis, indicating the absence of localized bound states. As $\eta$ increases ($> \eta_{c-}$), one localized bound state begins to emerge. Upon further increasing $\eta$ beyond a higher critical value ($> \eta_{c+}$), another localized bound state appears. 

In Fig.~\ref{UtdiffKap}, we firstly consider the case of $\kappa = 0$, where mode 1 and mode 2 are respectively coupled to the left and right environments, corresponding to two individual one-mode open systems coupled separately to their own environments. In this scenario, once $\eta > 0.2$, there are two localized bound states for mode 1 and mode 2 separately forming the localized bound state from the relatively strong coupling to the environment at the same frequency, namely $\omega_{l-}=\omega_{l+}$. Both poles located at the same point, as shown in the top panel of Fig.~\ref{UtdiffKap}. This behavior resembles that of a single-mode system, as discussed in previous research \cite{WMZ2012, Xiong2010}. When $\kappa > 0$, mode 1 and mode 2 couple with each other through the inter-mode coupling. The middle panel of Fig. \ref{UtdiffKap} shows 
for $\kappa = 0.25\omega_0$ that the low-frequency localized bound state ($\omega_{l-}$) appears for $\eta > 0.15$ (one pole) and the high-frequency localized bound state $\omega_{l+}$ appears for $\eta > 0.25$, see for example $\eta = 0.2$ and $\eta = 0.3$, there are one pole and two poles, respectively, as shown in the middle panel of Fig. \ref{UtdiffKap}. These two localized bound states lead to oscillatory dissipation dynamics with frequency $\omega_{loc} = \omega_{l_{-}} - \omega_{l_{+}}$, as illustrated in the next subsection. 

With further increasing the inter-mode coupling $\kappa$, we find that the low-freqenecy localized bound state occurs in the weaker coupling, and the high-frequency localized bound state occur for the stronger coupling. An example is shown in the bottom panel of Fig.~\ref{UtdiffKap}, where 
$\kappa = 0.75\omega_0$, one localized bound state emerges for the smaller coupling $\eta < 0.1$, but the second localized bound state  has not occurred even for the coupling is increased to $\eta = 0.3$. This is because a larger inter-mode coupling shifts the two eigenmode frequencies toward the opposite direction in the frequency domain, which affects the emergence of the localized bound states. As one can see from analytic expression Eq.~(\ref{LBS_lambda1}), the inter-mode coupling $\kappa$ and the detuning $\delta$ have the same contribution 
on shifting the localized bound state frequency when $\lambda=1$. One localized bound state is shifted to the higher frequency, while the other is shifted to the lower frequency so that the critical couplings $\eta_{c\pm}$ change significantly.

\begin{figure}[b]
    \centering
    \includegraphics[width=0.9\linewidth]{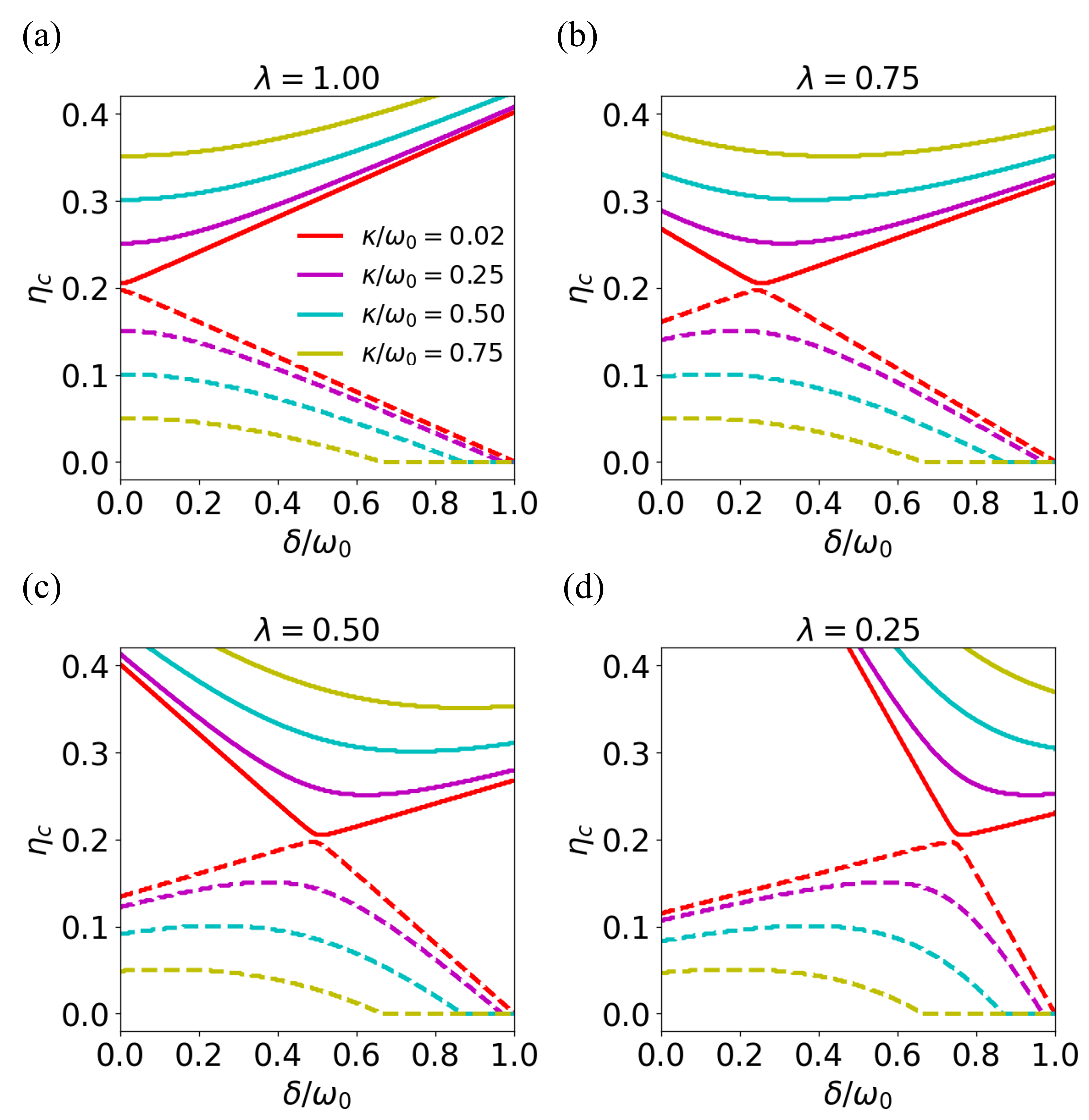}
    \caption{(Color online) The calculated critical system-environment coupling strength $\eta_{c_{-}}$(dashed line) and $\eta_{c_{+}}$(solid line) in various Ohmic environmental structure (a) $\lambda = 1$, (b) $\lambda = 0.75$, (c) $\lambda = 0.50$ and (d) $\lambda = 0.25$ by varying $\delta$ at different values of the inter-mode coupling $\kappa$. Here, we take $\omega_c = 5\omega_0$.}
    \label{JcvarLam}
\end{figure}

Figure \ref{JcvarLam} demonstrates the changes of the critical couplings $\eta_{c_{-}}$ and $\eta_{c_{+}}$ for the emergence of the 
two localized bound states characterized by the localized eigenmode frequencies $\omega_{l-}$ and $\omega_{l+}$, as the 
detuning $\delta$ varying  for several different inter-mode coupling $\kappa$.  
%is also supported by Fig. \ref{fig. JcvarLam_a}, where we numerically calculated the critical system-environment coupling strengths 
%$\eta_{c_{-}}$ and $\eta_{c_{+}}$ from Eq. (\ref{LBS}) in the Ohmic environmental structure with symmetric system-environment 
%coupling, as a function of the detuning $\delta$ with the inter-mode coupling $\kappa$. 
%In the following, we define two critical system-environment couplings, $\eta_{c+}$ and $\eta_{c-}$. 
When $\eta < \eta_{c-}$, no localized bound state exists; when $\eta_{c-} < \eta < \eta_{c+}$, one localized bound state occurs; and when $\eta > \eta_{c+}$, two localized bound states are presented.  As a result, the critical system-environment couplings $\eta_{c-}$ and $\eta_{c+}$ of the two localized bound states $\omega_{l-}$ and $\omega_{l+}$ decreases and  increases respectively as $\kappa$ or $\delta$ increases, see Fig.~\ref{JcvarLam}(a). 
When $\kappa/\omega_0 \sim 0$, e.g.,~$\kappa = 0.02\omega_0$, see the red lines in Fig.~\ref{JcvarLam}(a), the two-mode system behaves similarly as two independent single-mode systems coupled to their own environments \cite{WMZ2012, Xiong2010}. Furthermore, the critical system-environment coupling strength $\eta_{c_{-}} \!\rightarrow 0$ when $\kappa$ and $\delta$ become large, see for example $\kappa = 0.75 \omega_0$ and $\delta \sim 0.7\omega_0$, as shown in Fig.~\ref{JcvarLam}(a). In other words, the low-frequency localized bound state always exists regardless of the system-environment coupling strength under certain large values of $\kappa, \delta$. 

\begin{figure}[t]
    \centering
    \includegraphics[width=0.9\linewidth]{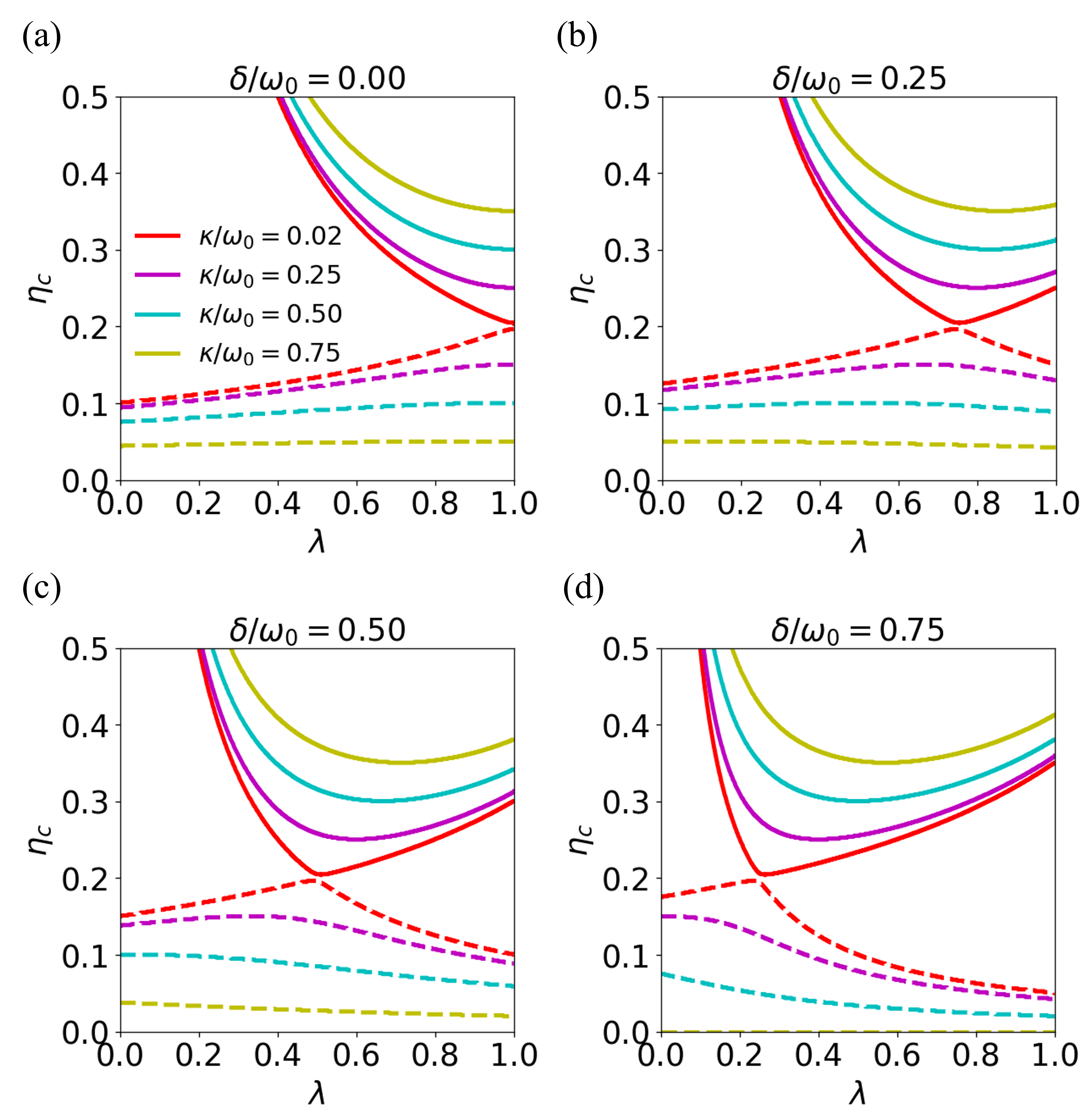}
    \caption{(Color online) The calculated critical system-environment coupling strength $\eta_{c_{-}}$(dashed line) and $\eta_{c_{+}}$(solid line) with the detuning (a) $\delta = 0.00\omega_0$, (b) $\delta = 0.25\omega_0$, (c) $\delta = 0.50\omega_0$ and (d) $\delta = 0.75\omega_0$ by varing $\lambda$ in Ohmic spectral density at different values of the inter-mode coupling $\kappa$. Here, we take $\omega_c = 5\omega_0$.}
    \label{JcvarDel}
\end{figure}

In Fig.~\ref{JcvarLam}, we also plot the critical system-environment coupling strengths $\eta_{c_{-}}$ and $\eta_{c_{+}}$ for the asymmetric couplings to the two environments, giving by the different values of $\lambda \neq 1 $. The case $\lambda=1$ is shown in Fig. \ref{JcvarLam}(a). Comparing with the symmetric couplings of $\lambda=1$, the plotted pattern is rotated clockwise as $\lambda$ decreases from $\lambda=1$, see Fig.~\ref{JcvarLam}(b)-(d). This is shown analytically by Eq.~(\ref{LBS}). Furthermore, in Fig.~\ref{JcvarDel}, we calculated the critical system-environment coupling strengths, $\eta_{c_{-}}$ and $\eta_{c_{+}}$ by varying $\lambda$ for different values of the detuning $\delta$ and the inter-mode coupling $\kappa$. The results are consistent with those presented in Fig.~\ref{JcvarLam}. In addition, for relatively strong inter-mode coupling, the environmental asymmetry on both sides does not affect the critical system-environment coupling strength $\eta_{c_{-}}$ (yellow dashed lines of Fig.~\ref{JcvarDel}). In fact, these illustrations are valid for all three types of Ohmic spectral densities given by Eq. (\ref{Ohmic-type}). 
%The results are consistent with our analytical calculation. \\

Moreover, compared to at most one localized bound state in a single-mode system for Ohmic-type spectral density \cite{WMZ2012, Xiong2010}, there are two localized bound states in a two-mode system due to the multiple pole conditions, see Eqs.~(\ref{LBS}) and (\ref{LBS_lambda1}). Therefore, we anticipate that there can be at most $2N$ localized bound states for an open quantum system, where $N$ is the number of band gaps in its structural or engineering environments, such as photonic crystals and semiconductors, etc. Additionally, the localized bound states emergence with one smaller system-environment coupling and one larger, determined by the frequency (or energy) shift from the inter-mode coupling and the detuning, akin to the construction and destruction of interference phenomena and manifest the multi-entanglement between the system and the environments.

\begin{figure*}[htb]
    \centering
    \includegraphics[width=0.9\textwidth]{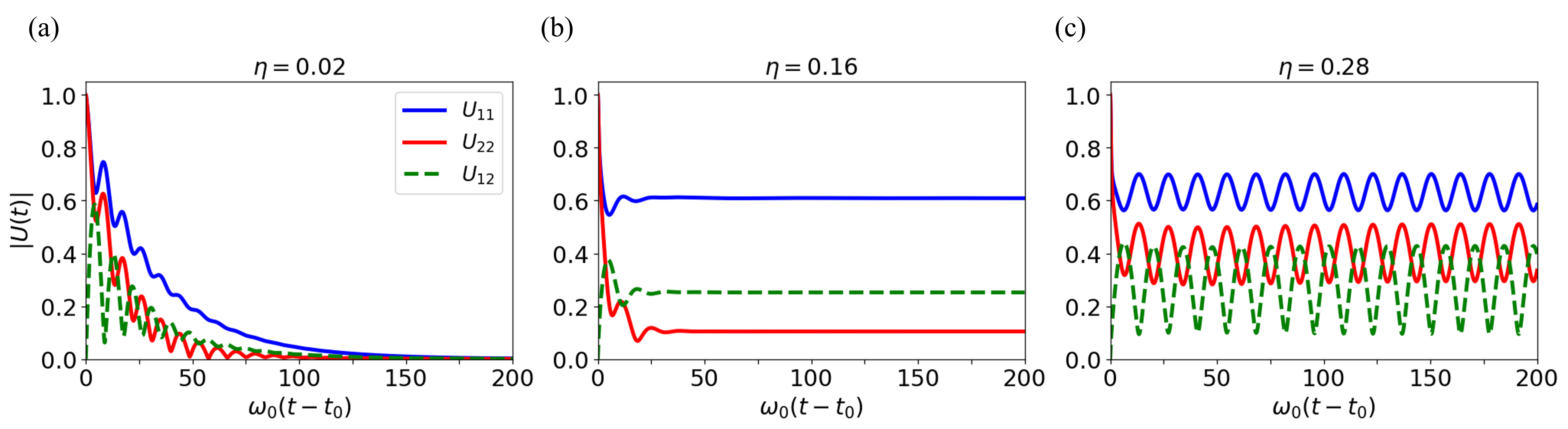}
    \caption{(Color online) The time evolution of the two-mode Green function $\abs{\bm{U}(t,t_0)}$ with different system-environment coupling strength $\eta$ with Ohmic spectral density ($s = 1$). The localized bound state appear when the system-environment coupling strength surpasses the critical point given by $\eta_{c_{\pm}}(s) = \omega_{\pm}/\omega_c\Gamma(s)$, which can be obtained from Eq.(\ref{LBS_lambda1}). Other parameters are taken as $\lambda = 1.00$, $\delta = 0.25\omega_{0}$, $\kappa = 0.25\omega_{0}$ and $\omega_c = 5\omega_0$. Note that $\eta_{c_{-}} \approx 0.13$ and $\eta_{c_{+}} \approx 0.27$ in this case.}
    \label{fig. U}
\end{figure*}

\subsection{Dissipationless dynamics of localized bound states}

The propagating function $\bm{U}(t,t_0)$ of Eq.~(\ref{analyticU}) characterizes the dissipation dynamics of open quantum system in general \cite{WMZ2012,WMZ2019}. More specifically, $\abs{{U}_{11}(t,t_0)}$ and $\abs{{U}_{22}(t,t_0)}$ give the amplitudes of the mode 1 and mode 2 respectively, and $\abs{{U}_{12}(t,t_0)}$ describes the coherence between the two modes. In the weak coupling $\eta < \eta_{c_{-}}$, no localized bound state emerges, the amplitudes of the two modes gradually decay to zero with a short-time oscillation through the energy exchanges with its environment, as shown by the blue and red solid lines in Fig.~\ref{fig. U}(a). The coherence dynamics between the two modes also 
decays to zero as a decoherence effect, see the green line in Fig.~\ref{fig. U}(a). The short-time oscillation is the short-time non-Markovian memory dynamics representing the backactions between the system and environments due to the system-environment interactions. 
%Note that, the environmental back-action can also induce the information exchange between the mode 1 and mode 2 through 
% the inter-mode coupling $\kappa$, which can be seen clearly in the dynamics of $\abs{\bm{U}_{12}}$ [green dashed line of Fig. \ref{fig. U_a}].

When $\eta_{c_{+}} > \eta \geq \eta_{c_{-}}$, one localized bound state with the low frequency $\omega_{l-}$ occurs, 
leading to a dissipationless stationary state, see Fig. \ref{fig. U}(b), as a long-time non-Markovian memory effect. 
In Fig. \ref{fig. U}(b), both $\abs{{U}_{11}(t,t_0)}$ and $\abs{{U}_{22}(t,t_0)}$ approach to a constant, because the inter-mode 
coupling $\kappa$ mixes the two modes with the environments in the formation of the localized bound state $\omega_{l-}$. 
In other words, both $\abs{{U}_{11}(t,t_0)}$  and $\abs{{U}_{22}(t,t_0)}$ contains the contribution from the 
low-frequency localized bound state. The inter-mode 
coupling $\kappa$ also induce the dissipationless coherence between the original two modes, even though there occurs only one localized bound state. This is shown by the green line in Fig. \ref{fig. U}(b), a non-zero constant value of $\abs{{U}_{12}(t,t_0)}$ in the steady state.  
As $\eta$ increasing further, the amplitude of the localized bound state increases \cite{wu2010non}, manifesting the 
stronger non-Markovian memory effects. When $\eta \geq \eta_{c_{+}}$, the high-frequency localized bound state also emerges. The frequency difference of the two localized bound states causes a dissipationless stationary oscillation, as a dissipationless coherence dynamics between the two localized bound states, as shown in Fig.~\ref{fig. U}(c). This dissipationless coherence dynamics fully immunes the usual environmental-induced decoherence, because the localized bound states themselves 
are generated from the coupling between the system and the environments \cite{WMZ2012,WMZ2019}.
%In the long-time limit $t = t_s \rightarrow \infty$, the system is dissipationless when localized bound state exists, 
%which is only the dissipationless localized bound states terms of Eq. (\ref{analyticU}) remain, 

\begin{figure*}[!htb]
    \centering
    \includegraphics[width=0.9\linewidth]{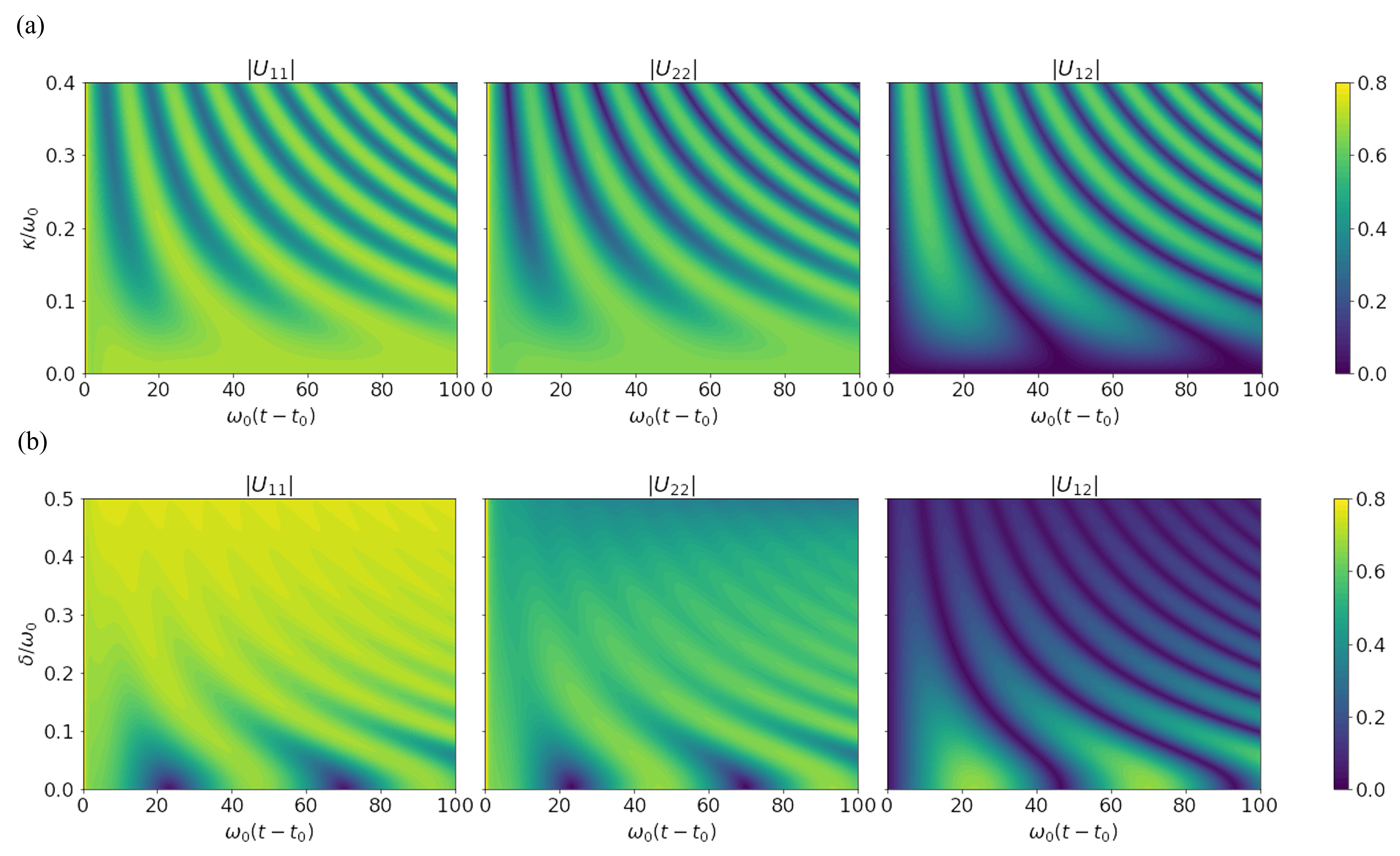}
    \caption{(Color online) Manipulation of the dissipationless system occupations and coherence in a two-mode open system. Other parameters are taken as (a) $\lambda = 1.00$, $\delta = 0.10\omega_0$, $\eta = 0.30$, $\omega_c = 5\omega_0$, and (b) $\lambda = 1.00$, $\kappa = 0.10\omega_0$, $\eta = 0.30$, $\omega_c = 5\omega_0$.}
    \label{fig. mnpU}
\end{figure*}

In Fig.~\ref{fig. mnpU}, we present the contour plots to demonstrate the dissipationless dynamics of the localized bound states for the continuous variation of the inter-mode coupling $\kappa$ and the detuning $\delta$. In order to clarify their influence to localized bound state, we set a strong system-environment coupling as $\eta = 0.3$. In the strong coupling regime, not only the effect of localized bound state is more crucial, but also the system can reach the steady-state limit more rapidly ($\omega_0 t_s \geq 20$, for $\eta = 0.28$), as shown in Fig.~\ref{fig. U}. Therefore, Fig.~\ref{fig. mnpU} can be understood better by only looking at the amplitudes of the localized modes and the interference (coherence) between them in the Green function $U(t,t_0)$, which are given by the first term in Eq.~(\ref{analyticU}),
%\begin{widetext}
    \begin{align}
      \bm{U}_l(t, t_0) \equiv & ~\bm{\mathcal{Z}}(\omega_{l-})e^{-i\omega_{l-}(t - t_0)}
      + \bm{\mathcal{Z}}(\omega_{l+})e^{-i\omega_{l+}(t - t_0)} .
%       & \abs{U_{ij}(t_s-t_0)} = \sqrt{\mathcal{Z}_{ij}^2(\omega_{l-}) + \mathcal{Z}_{ij}^2(\omega_{l+}) + 2\mathcal{Z}_{ij}%^2(\omega_{l-})\mathcal{Z}_{ij}^2(\omega_{l+})\cos{\left[(\omega_{l-} - \omega_{l+})(t_s - t_0)\right]}}
        \label{U_st} 
    \end{align}
%\end{widetext}
The second term in Eq.~(\ref{analyticU}) describes the dissipation dynamics which vanishes whatever in the steady-state limit $t \rightarrow t_s$. As a result,  the diagonal matrix element of $\abs{\bm{U}_l(t-t_0)}$ defined in Eq.~(\ref{U_st}) identifies the amplitudes of the two localized bound states and the coherence between them, which are strongly related to the long-time non-Markovian effects induced by the back-action from the environment. Specifically, the diagonal matrix element  $\abs{{U}_{l,ii}(t-t_0)}$ characterizes the rest of the initial information of mode $i$ in the decoherence dynamics. 
%That is to say, a large amount of $\abs{\bm{U}_{l,ii}(t-t_0)}$ counts the undissipated part of the mode $i$. 
Moreover, the decoherence dynamics do not influence the information exchange between the undissipated two modes through the system-environment coupling. It results in the formation of dissipationless coherence between the two modes during their evolution, represented as the mutual entanglement between the two modes. This is characterized by the off-diagonal matrix elements $\abs{{U}_{l,12}(t-t_0)}$ and $\abs{{U}_{l,21}(t-t_0)}$. 
As we can see, when the inter-mode coupling $\kappa$ gets stronger, the amplitude of $\abs{U_{12}(t,t_0)}$ increases, as shown in Fig.~\ref{fig. mnpU}(a). In other words, one can use the inter-mode coupling to control the dissipationless coherence between the two modes. On the contrary, increasing the detuning $\delta$ leads to a reduction in the coherence of the two modes,  as shown in Fig.~\ref{fig. mnpU}(b).  Thus, one can not only control the amplitudes of the localized bound states but also manipulate the oscillation frequency between them by using the inter-mode coupling $\kappa$ and the detuning $\delta$,  as one can see clearly in Fig. \ref{fig. mnpU}.
%This indicates that the energy level difference between the two modes is detrimental to their coherence, 
%thereby reducing the mutual memory effect.
% Also note that as shown in Eq. (\ref{U_st}), when there exist two localized bound states, the steady-state limits of system 
%occupations $\abs{U_{11}}$ and $\abs{U_{22}}$ as well as the system coherence $\abs{U_{12}}$ oscillate at a frequency 
%${\omega_{l_-} - \omega_{l_+}}$ that is controlled by $\kappa$ and $\delta$. As a result, both the system occupations 
%$\abs{U_{11}}$ and $\abs{U_{22}}$, and the system coherence $\abs{U_{12}}$ can be manipulated by tuning 
%the inter-mode coupling $\kappa$ or the detuning $\delta$.

%Both the system occupations $\abs{U_{11}}$ and $\abs{U_{22}}$, and the system coherence $\abs{U_{12}}$ can be manipulated by tuning the inter-mode coupling $\kappa$ or the detuning $\delta$, see Fig. \ref{fig. mnpU}. Specifically, when there exist two localized bound states, the system occupations as well as the system coherence oscillate at a frequency ${\omega_{l_-} - \omega_{l_+}}$, see Eq. (\ref{U_st}), which can be controlled by tuning the inter-mode coupling $\kappa$ and detuning $\delta$, as shown in Fig. \ref{fig. mnpU}. {\color{blue}Replenish amplitude and memory by the inter-mode coupling and the detuning..., see Fig. \ref{fig. mnpU}}

\section{Conclusion}
In this work, we explored the dissipationless localized bound state dynamics of a coupled two-mode open quantum system 
with our exact master equation theory.  We show that the dissipationless localized bound states, as the long-time non-Markovian phenomenon, is robust against decoherence. 
We identify the conditions for the emergence of the localized bound states in this open system with the Ohmic-type spectral densities. The corresponding critical system-environment couplings are determined and studied under different values of the 
inter-mode couplings and detuning. We also show that the dissipationless localized bound state dynamics and the associated coherence can be manipulated through the inter-mode coupling and the detuning between the two modes. 
The analytical approach for the dissipationless localized bound states considered in this work is indeed universal. It can be 
further applied to arbitrary many-body systems because all the dissipationless localized bound states are solved from the non-equilibrium Green functions which are valid for any interacting many-body systems. 
The robustness of their dissipationless against decoherence indicates that the localized bound states can be a promising 
candidate for a practical realization of decoherence-free qubit manipulations.

In fact, the main obstacle in the development of the quantum technology is the decoherence induced by the inevitable couplings between quantum devices and their environments. To overcome the problem of decoherence, alternatively a lots of research have been focused on topological quantum computation in which topological qubits, known as Majorana zero modes (MZMs), have been considered to be immune to local perturbations and thereby robust against decoherence. However, the search of the Majorana bound states has not been successful in experiments. Meanwhile, 
theoretically we have proven that Majorana zero mode bound states (if exist) still suffer from the same decoherence problem as other qubit systems  \cite{Lai2018,Yao2020,YW2020,Yao2023}.     
In this work, although we only dealt with a simple open quantum system, the properties of localized mode bound states are 
indeed general to other open quantum systems, as we have pointed out. In particular, for structural and engineering 
environments, such as photonic and electronic crystal materials with multiband band structures, each band bap can host 
one or more localized bound states.  
Therefore, as a straightforward extension of semiconductor materials utilized in the traditional computing industry, the concept of localized 
bound states lied on band gap structures of photonic and electronic materials and the manipulation of quantum coherence with localized 
bound states may provide a new avenue for the quantum computation and quantum technology development.

\iffalse
Moreover, quantum coherence has been seen as a mechanism that nature uses to achieve efficient energy transfer. This phenomenon is widely observed in femtosecond multidimensional spectroscopy of several pigment-protein complexes, including the Fenna-Matthews-Olson complex and the light-harvesting complex II \cite{Engel2007, Lambert2013, Cao2020}. It is often regarded as an experimental signature of nontrivial quantum effects in light harvesting, characterized by oscillatory signals indicating quantum coherence in biological systems. Instead of avoiding decoherence entirely, nature utilizes site energies and inter-mode coupling to facilitate energy transport. Considering the possibility of dissipationless long-term non-Markovian memories, multi-mode localized bound states may provide insights into how biological systems resist decoherence and how to develop the new quantum devices of energy storages.  We leave these for future research.
\fi

\begin{acknowledgments}
This work is supported by National Science and Technology Council of Taiwan, Republic of China, under Contract
No. MOST-111-2811-M-006-014-MY3.
\end{acknowledgments}

% If you have acknowledgments, this puts in the proper section head.
%\begin{acknowledgments}
% put your acknowledgments here.
%\end{acknowledgments}

% The \nocite command causes all entries in a bibliography to be printed out
% whether or not they are actually referenced in the text. This is appropriate
% for the sample file to show the different styles of references, but authors
% most likely will not want to use it.
\nocite{*}

% \bibliography{TwoModeRef}% Produces the bibliography via BibTeX.

\end{document}